%% file: main.tex
\documentclass[sigconf]{acmart}
\def\BibTeX{{\rm B\kern-.05em{\sc i\kern-.025em b}\kern-.08emT\kern-.1667em\lower.7ex\hbox{E}\kern-.125emX}}
    
\usepackage{nicefrac}
\usepackage{wasysym}
\usepackage{siunitx}
\usepackage{array,framed}
\usepackage{booktabs}
\usepackage{
  color,
  float,
  epsfig,
  wrapfig,
  graphics,
  graphicx,
  subcaption
}
\usepackage{textcomp}
\usepackage{setspace}
\usepackage{latexsym,fancyhdr,url}
\usepackage{enumerate}
\usepackage{algorithm2e}
\usepackage{algpseudocode}
\usepackage{graphics}
\usepackage{xparse} 
\usepackage{xspace}
\usepackage{multirow}
\usepackage{csvsimple}
\usepackage{balance}

\usepackage{
  tikz,
  pgfplots,
  pgfplotstable
}
\usepackage{hyperref}

\usetikzlibrary{
  shapes.geometric,
  arrows,
  external,
  pgfplots.groupplots,
  matrix
}

\pgfplotsset{compat=1.9}

\usepackage{mathtools,}

\DeclareMathAlphabet{\mathcal}{OMS}{cmsy}{m}{n}

\DeclareGraphicsExtensions{%
    .png,.PNG,%
    .pdf,.PDF,%
    .jpg,.mps,.jpeg,.jbig2,.jb2,.JPG,.JPEG,.JBIG2,.JB2}

\input{defs}
 \setcopyright{none}
 \renewcommand\footnotetextcopyrightpermission[1]{}

\begin{document}
\fancyhead{}
\def\thetitle{Gender and Prestige Bias in Coronavirus News Reporting}
\title{\thetitle}

\author{Rebecca Dorn}
\affiliation{\small{USC Information Science Institute}
\\Marina del Rey, CA, USA\\rdorn@usc.edu}

\author{Yiwen Ma}
\affiliation{\small{USC Information Science Institute}
\\Marina del Rey, CA, USA\\yiwenma@usc.edu}

\author{Fred Morstatter}
\affiliation{\small{USC Information Science Institute}
\\Marina del Rey, CA, USA\\fredmors@isi.edu}

\author{Kristina Lerman}
\affiliation{\small{USC Information Science Institute}
\\Marina del Rey, CA, USA\\lerman@isi.edu}

\date{}

\input{abstract}
\maketitle

\section*{CCS Concepts}
Information systems $\rightarrow$ Data mining, Social networks;
Computing methodologies $\rightarrow$ Natural language processing

\section*{Keywords}
gender bias; prestige bias; ideological bias; news reporting; expert sources; named entity recognition; dependency parsing

\input{intro}    
\input{background}
\input{methodology}
\input{results}
\input{conclusion}

\section*{Ethics Statement}
\input{ethics}
\section*{Acknowledgements} This work was supported, in part, by the Defense Advanced Research Projects Agency  under contract W911NF192027.
\bibliographystyle{ACM-Reference-Format}
\bibliography{bib}


\end{document}

%% file: defs.tex
\usepackage{xparse}
\newcommand{\bnm}{\begin{newmath}}
\newcommand{\enm}{\end{newmath}}

\newcommand{\bea}{\begin{eqnarray*}}%
\newcommand{\eea}{\end{eqnarray*}}%

\newcommand{\bne}{\begin{newequation}}
\newcommand{\ene}{\end{newequation}}

\newcommand{\bal}{\begin{newalign}}
\newcommand{\eal}{\end{newalign}}

\newenvironment{newalign}{\begin{align}%
\setlength{\abovedisplayskip}{4pt}%
\setlength{\belowdisplayskip}{4pt}%
\setlength{\abovedisplayshortskip}{6pt}%
\setlength{\belowdisplayshortskip}{6pt} }{\end{align}}

\newenvironment{newmath}{\begin{displaymath}%
\setlength{\abovedisplayskip}{4pt}%
\setlength{\belowdisplayskip}{4pt}%
\setlength{\abovedisplayshortskip}{6pt}%
\setlength{\belowdisplayshortskip}{6pt} }{\end{displaymath}}

\newenvironment{newequation}{\begin{equation}%
\setlength{\abovedisplayskip}{4pt}%
\setlength{\belowdisplayskip}{4pt}%
\setlength{\abovedisplayshortskip}{6pt}%
\setlength{\belowdisplayshortskip}{6pt} }{\end{equation}}

\newcounter{ctr}

\newcounter{mytable}
\def\mytable{\begin{centering}\refstepcounter{mytable}}
\def\endmytable{\end{centering}}

\newcounter{myfig}
\def\myfig{\begin{centering}\refstepcounter{myfig}}
\def\endmyfig{\end{centering}}

\newlength{\saveparindent}
\newlength{\saveparskip}
\newcommand{\E}{{\rm I\kern-.3em E}}

\renewcommand{\eqref}[1]{\mbox{Equation~(\ref{#1})}}

\def \part {part}

\renewcommand{\paragraph}[1]{\vspace*{6pt}\noindent\textbf{#1}\;}

\def \blackslug{\hbox{\hskip 1pt \vrule width 4pt height 8pt
    depth 1.5pt \hskip 1pt}}
\def \qed{\quad\blackslug\lower 8.5pt\null\par}

\newcounter{mynote}[section]

\newcommand\ignore[1]{}

\newcounter{rcnote}[section]

\newcounter{mrnote}[section]

\newcounter{fknote}[section]

\newcounter{anote}[section]

\DeclareMathSymbol{\mlq}{\mathord}{operators}{``}
\DeclareMathSymbol{\mrq}{\mathord}{operators}{`'}

\newcommand{\rhf}[2]{R_{f, \gamma}}

\DeclareDocumentCommand{\edist}{o o}{
  \ensuremath{
    \IfNoValueTF{#1}{{d}}{{\sf d}(#1,#2)}
  }
}

\newcommand{\mathcmd}[1]{\ensuremath{#1}\xspace} 

\newcommand{\s}{\mathcmd{s}}

\newcommand{\olrk}[1]{\ifx\nursymbol#1\else\!\!\mskip4.5mu plus 0.5mu\left(\mskip0.5mu plus0.5mu #1\mskip1.5mu plus0.5mu \right)\fi}

\NewDocumentCommand{\indseq}{ O{1} O{r} }{{#1}\ldots {#2}}


%% file: abstract.tex
\begin{abstract}

Journalists play a vital role in surfacing issues of societal importance, but their choices of what to highlight and who to interview are influenced by societal biases. In this work, we use natural language processing tools to measure these biases in a large corpus of news articles about the Covid-19 pandemic. Specifically, we identify when experts are quoted in news and extract their names and institutional affiliations. We enrich the data by classifying each expert's gender, the type of organization they belong to, and for academic institutions, their ranking. Our analysis reveals disparities in the representation of experts in news. We find a substantial gender gap, where men are quoted three times more than women. The gender gap varies by partisanship of the news source, with conservative media exhibiting greater gender bias. We also identify academic prestige bias, where journalists turn to experts from highly-ranked academic institutions more than experts from less prestigious institutions, even if the latter group has more public health expertise. Liberal news sources exhibit slightly more prestige bias than conservative sources. Equality of representation is essential to enable voices from all groups to be heard. By auditing bias, our methods  help identify blind spots in news coverage.

\end{abstract}

%% file: intro.tex
\section{Introduction}
In times of crisis people turn to news media for information and to make sense of the world; journalists, in turn, seek out experts and opinion leaders to interview and then help communicate their knowledge to the public. Mass media does not simply convey  information to the public but also shapes what is seen and what is deemed important~\cite{mccombs1972agenda}. The interplay between mass media and the public creates a cycle that amplifies attention to concerns and influences public policy. Given the media's role in identifying issues of societal importance, it is therefore critical that it equitably reflects the interests of all stakeholders.

Representation of groups and individual social identity in the media is one of the fundamental questions of equity. Does the media adequately represent issues that are important to women, ethnic minorities, the elderly, and the disadvantaged? Does it capture the lived experience of these groups, the challenges they face? Or does it focus on the concerns of the privileged few? One mechanism for improving equity is to ensure that the pool of journalists and reporters reflects society's diversity. 
However, journalists are predominantly men and often choose to interview subjects whose gender identity matches their own \cite{manoso2019gender}. 

Another mechanism to improve equity is to diversify the pool of subjects that journalists pay attention to. 
For example, by talking to women, journalists will surface their views and concerns. This is important, because women typically bear a larger share of care responsibilities, and their concerns may bring up issues with childcare, for instance, that may not be visible to men. 
Moreover, if journalists solely focus on sources from the same few prestigious academic institutions, they lose the geographic and socio-economic diversity that comes from interviewing experts from a range of institutions. This may introduce additional blind spots in news coverage.

Auditing gender representation in the news---or the representation of other identities---has proven difficult due to the challenges of extracting representations from the text of the news stories.
Previous studies have identified gender bias in news reporting~\cite{shor2015paper}; however, they have generally relied on manually curated data or were limited to certain media types, and thus do not scale to the size of the media ecosystem. Addressing the question of bias in the news media at scale calls for automated methods.
In this study we use natural language processing (NLP) methods to automate media analysis, which enables us to scale our bias audit of news across longer time periods and across more media sources. We focus on gender and academic prestige bias in the coverage of the Covid-19 pandemic. 
When the novel coronavirus emerged, little was known about the severity of the disease it caused, what mitigations were effective and their benefits and costs. As researchers learned more about the disease,  public officials used these findings as a basis for policy recommendations. Journalist sought out experts from the research community and government agencies to communicate the research findings,  policy recommendations, and their trade-offs to the public.
We analyze thousands of news stories from six popular media sources along the breadth of US political spectrum to identify the experts the journalists turned to. We analyze three left leaning news sources and three right leaning sources to enable analysis by partisan bias and accommodate a variety of linguistic styles.


Our analysis reveals a gender gap in news coverage where women appear much less frequently among the experts quoted by journalists than men. The gender gap varies by political ideology of the news source, with liberal media coming closer to gender parity than conservative media. In addition to gender, we look at the institutional affiliations of the experts and classify their academic prestige. We identify prestige bias, in which experts from the higher-ranked academic institutions are quoted more frequently than experts with less prestigious affiliations. We find that prestige bias varies slightly by ideology of the reporting source.

One possible explanation for the observed bias is that women are a minority in science and medicine. However, women make up the majority of doctoral students and junior faculty in public health and biomedical sciences
~\cite{schisterman2017changing}, both of which are fields relevant to the Covid-19 pandemic. Graduate-level public health degrees have been awarded to more women than men since 1979, with 73\% of such degrees awarded to women in 2017 \cite{leider2018trends}. Therefore, the gender disparity we observe is likely not due to a shortage of experts but due to individual biases of reporters and media sources.



Our analysis of the gender and prestige of experts quoted in the news during the Covid-19 pandemic  answers the following research questions:
\begin{itemize}
\item \textbf{Gender Bias:} Are women underrepresented among experts whom journalists turn to for information about the  pandemic?
\item \textbf{Ideological Gender Bias: } Does the gender gap vary by ideological leaning of news source?
\item \textbf{Prestige Bias: } Is there media preference for experts from highly ranked institutions? 
\item \textbf{Ideological Prestige Bias:} Does the prestige gap change with political leaning of news outlet?
\end{itemize} 


%% file: background.tex
\section{Related Work}
There has been work analyzing the gender composition of experts in television news. Scott et al. discovered that from September 25 to October 6, 2006 and May 14 to May 25, 2007, 14.7\% of people featured in PBS NewsHour were women \cite{scott2010fair}. 
The authors also found that 13.7\% of experts had academic affiliations, 4.3\% from think tanks and 42.9\% with governmental affiliations.

The role of gender in international news media use of non-coronavirus specific experts has been documented. Niemi et al. found that less than 30\% of experts interviewed in Finnish news journalism are women \cite{niemi2017gendered}. Lidia Ma{\~n}oso Pacheco found a high correlation between journalist and subject gender in 68 British and English newspaper articles \cite{manoso2019gender}. Kitzinger et al. analyzed 51 in-depth profiles of men and women scientists and found that 5 men are used for every 1 woman scientist \cite{kitzinger2008gender}.

Only manual analyses of American Coronavirus news experts exist. Fletcher et al. \cite{fletcher2021gender} reviewed a total of 4,463 articles from 9 U.S. news sources dating April 1, 2020 to April 15, 2020 and found 35.9\% of the 2,297 experts were women. In a special report from Luba Kassova that looked at the frequency of men and women in 2,100 quotes between March 1, 2020 and April 15, 2020, men were quoted three times as much as women \cite{kassova2020missing}. Kassova additionally found that women are less likely to be protagonists in news stories and more likely to provide subjective views over expertise.

Large scale analysis of North American news experts exist, though not specific to Coronavirus. 
Asr. et al. introduced a tool for large scale gender analysis in news quotes in The Gender Gap Tracker \cite{asr2021gender}, which takes a sentence and returns people quoted and mentioned with their inferred gender identities. Methods of extraction include syntactic, heuristic and floating quote approaches. The software is illustrated on seven Canadian news outlets, where the authors found that men are represented three times as much as women from October 2018 to September 2020.

Large-scale tools have been used to analyze the difference in how men and women are featured in the news.
LDA topic modelling is performed on two years worth of American and Canadian news articles by Rao et al. \cite{rao2021gender}. Persons quoted and their genders are gathered using The Gender Gap Tracker. 
Contrary to our results, the authors found that women are more represented in articles related to healthcare. 
An analysis of gender, fame and sentiment is done by Shor et al. \cite{shor2022women}. The dataset used combines 14 million persons mentioned throughout 1,323 news outlets with a manual analysis of select Wikipedia pages. The authors looked at sentiment scores for adjectives used with each person, and found that as women become more famous the media attention recieved becomes increasingly negative. Separately, Shor et al. analyzed gender and public interest while controlling for occupation and age \cite{shor2019large}. The authors looked at over 20,000 persons from over 2,000 news sources. They found that when men and women have similar occupations and ages, women obtain higher public interest but less media coverage.

One of the most frequently observed forms of social learning is where people observe and mimic seemingly competent and therefore admirable individuals \cite{jimenez2019prestige}. Jimenez et al. explained how first order cues of prestige (initially observable traits) are used to assume prestige when quality information is lacking, though these cues may be wrong and deceptive \cite{jimenez2019prestige}. Additionally, upward mobility in academia is limited. In a survey of n = 348 universities, 20\% of faculty positions are inhabited by 8 universities \cite{wapman2022quantifying}. The same survey found that only 5\% to 23\% of faculty members from United States universities hold doctorates from less prestigious institutions, and that 64\% of Universities have no departments listed as top 10 \cite{wapman2022quantifying}. 




%% file: methodology.tex
\section{METHODS}

\subsection{Data}
The AYLIEN Coronavirus Dataset consists of 1,673,353 news articles related to the Coronavirus pandemic collected from over 440 international news sources. This data is aggregated, analyzed, and enriched by AYLIEN using AYLIEN's News Intelligence Platform\footnote{https://aylien.com/resources/datasets/coronavirus-dataset}. We use the article attributes raw article text, article title, news source name, and publication date and time. We analyze AYLIEN Coronavirus related news articles from six US-based news sources: Huffington Post (HUFF), Cable News Network (CNN), The New York Times (NYT), The New York Post (NYP), Fox News (FOX), and Breitbart News Network (BREIT) between January 6, 2020, and July 31, 2020. These six news outlets are chosen because they collectively exemplify an ideological spectrum in news reporting while all having some partisan bias. This allows us to separate news outlets into two distinct groups. Additionally, having 6 news outlets ensures we cover a variety of linguistic style. This subset totals 66,368 articles: 9,897 articles from the New York Times, 17,765 from CNN, 19,911 from Fox News, 7,609 from Breitbart, 13,391 from New York Post and 6,625 from the Huffington Post.

\subsection{Expert Quote Extraction}
Fig. \ref{fig:examplequote} shows an example of how journalists quote experts using three different sentence structures. The components of interest are reported speech, reported verb, person and organization. Reported speech (RSPEECH) directly quotes or indirectly reconstructs the words of the speaker. A reporting verb (RVERB) is used to introduce or conclude reported speech (e.g. ``report'', ``acclaim'', ``told''). The person is the speaker being quoted. An organization is the institution associated with the speaker. We consider expert quotes to be any permutation of these components. We find sentences quoting experts by taking the union of two approaches:

\begin{figure*}[h!]
\includegraphics[width=\textwidth]{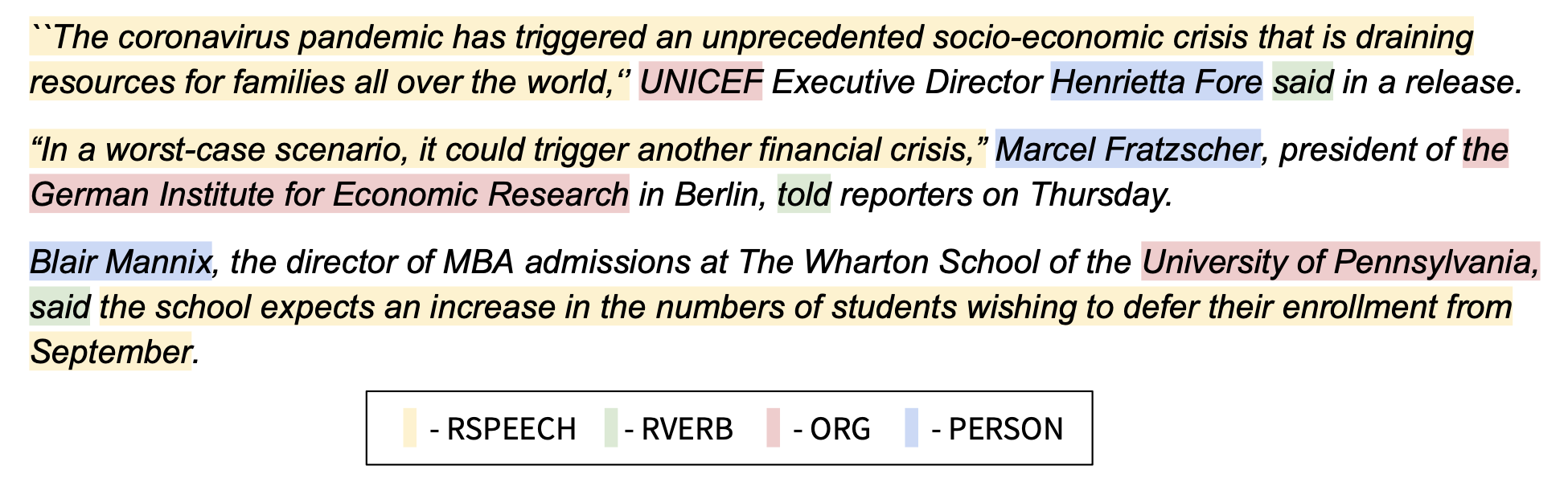}
\caption{\label{fig:examplequote} \textbf{Examples of Expert Quotes.} Examples capture three varieties of quote structure. RSPEECH (Reported Speech) is the portion of the quote containing an exact quote or reconstruction of what the speaker previously said. RVERB (Reporting Verb) refers to the verb introducing or concluding reported speech ("say", "said", "explains", etc.). PERSON refers to the speaker of the (reported) quote. ORG refers to the organization affiliated with the speaker. Quotes are considered expert quotes if it has the presence of RSPEECH, RVERB, PERSON and ORG. We consider a sentence as containing both RSPEECH and RVERB if it contains one of 262 Reporting Verbs, as a Reporting Verb implies the presence of Reported Speech. We use Named Entity Recognition (NER) to determine whether a sentence features a PERSON and ORG.}
\end{figure*}


\subsubsection{Named Entity Recognition (NER)}
The three most common reporting verbs are ``said'', ``say'' and ``says''. The most common pattern quoting experts is:
 \begin{quote}
     ``$[RSPEECH]$," (``said"$|$``say"$|$``says") [PERSON]
 \end{quote}
Where $|$ denotes logical \textit{or} and [PERSON] denotes speaker. 
This pattern is captured using the following regular expression:
 \begin{quote}
     ``$\s$([a-zA-z0-9?',.$\s$()])*\s,"(said$|$say$|$says)([a$-$zA$-$z0$-$9?',\s()])*
 \end{quote}
The NLP library SpaCy offers an NER library pretrained on web text with entity labels including person, organization, date and location \cite{spacy2}. We use SpaCy's NER on sentences following this pattern and look for PERSON entities listed outside of quotation marks.

\subsubsection{The Gender Gap Tracker} The second method we use to find speakers is that of The Gender Gap Tracker Project~\cite{asr2021gender}. The syntactic method from The Gender Gap Tracker identifies quotes following a clausal complement structure, where a dependent verb is featured with an internal subject. Sentences following this structure are only kept if they feature one of 262 reporting verbs. The second Gender Gap Tracker method we utilize identifies reported speech introduced directly before or after the reporting verb ``according to.'' Due to the difficulty in finding affiliated organizations, we choose to omit the floating quote method which finds sentences where reported speech takes a full sentence and the speaker is introduced elsewhere.

When an expert is quoted in a news article, the journalist typically introduces the expert, specifying their position and affiliation. To help focus our data collection only on expert speakers, we require speakers to be present alongside an organizational affiliation. On all sentences collected, we run NER and retain only those sentences where NER identifies an organization (ORG entity).

\subsection{Classifying Gender}
The Python library gender-guesser implements a gender prediction program built on a database of 45,376 names with each name's most likely gender identity \cite{genderguesser}. The possible gender predictions for a single person are ``male", ``female", ``andy" (androgynous) and ``unknown". For each person quoted, we run gender-guesser on the first string before a space (i.e., first name) to obtain that name's most common gender association \cite{santamaria2018comparison}. 

The gender labels include ``male" and ``female" though would be more accurately described as man/masculine and woman/feminine.  We acknowledge that gender is non-binary and not captured by a person's first name. Classifying by common gender affiliation with names captures reader perception of gender, not the expert speakers’ actual gender identification. The discussion section further elaborates on the inability of a single androgynous category to adequately capture non-binary non-cisgender gender identities.

\subsection{Classifying Organization Prestige}
During the Covid-19 pandemic, scientists, epidemiologists, and public health experts from a variety of different organizations worked to define our understanding of the disease and to define public policy. These experts came from academic institutions (e.g., Brown University), federal bodies (e.g., the Centers for Disease Control and Prevention), and a variety of think tanks (e.g., the Hoover Institution). Journalists turned to these experts for information and guidance to share with the public.

We use fuzzy string matching, a mechanism that generates similarity scores between two strings, to determine whether organization affiliations reference academic institutions, federal bodies, or think tanks. For example, fuzzy string matching would find that ``The University of Maryland - College Park" matches to ``The University of Maryland" with a score of 90. Journalists typically introduce organizations with their full names, thus we do not accomodate for organization abbreviations.


\subsubsection{Academic Institutions} We use Times Higher Educations' 2015 World University Rankings\footnote{https://www.timeshighereducation.com/world-university-rankings/2016/world-ranking/methodology}. This list gives 400 University names as well as their ranking. Rankings are determined by factors including teaching, research, citations, industry income, and international outlook.

\subsubsection{Federal Bodies} We compile a list of Federal Bodies by web scraping the U.S. Government Services and Information's index of Federal Departments and Agencies\footnote{https://www.usa.gov/federal-agencies}. This list includes only federal agencies therefore nothing at the state level.

\subsubsection{Think Tanks} One of the most popular think tank definitions is by McGann and Weaver: ``non-governmental, not-for-profit research organisations with substantial organisational autonomy from government and from societal interests such as firms, interest groups, and political parties'' \cite{weaver2017think, pautz2011revisiting}. Think tanks frequently focus on public policy. We use the open source database On Think Tanks\footnote{https://onthinktanks.org/open-think-tank-directory/}, which includes over 3,200 global think tanks and provides fields including region, topic, website and office address.

For each sentence, we measure similarity between NER-identified organization and organization names listed in these databases. We manually review a sample of NER-extracted organizations, the organization name most closely matching and the distance metric calculated for the two strings. For all three databases, we consider a match if the similarity score is greater than or equal to 90. To minimize noise, organizations consisting of two or fewer characters in the name are ignored. We sample 25 random organizations of two or fewer characters to ensure minimal impact. We find that the most common two-character string is ``''s", followed closely by strings ``'m" and ``AP".

%% file: results.tex
\section{RESULTS}

We extract 89,130 expert sources (pairs of speakers and their affiliated organizations): 19,137 pairs from HUFF, 17,156 from CNN, 18,828 from NYT, 4,129 from NYP, 22,226 from FOX and 7,654 from BREIT. The Gender Gap Tracker accounts for 26.7\% of these extractions, and Named Entity Recognition-based for the rest. Our methods improve the number of extractions by 65,263 pairs. The scale increase from adding our method helps promote accuracy and efficiency in studies of inequality.

For precision evaluation, we run our method on 100 randomly sampled articles and manually annotate each extraction. Extractions are labeled correct if they contain RSPEECH from a PERSON with an ORG affiliation. The precision from this sample is 64.7\%. The method most commonly fails for instances where the ORG is the news outlet rather than a professional affiliation. For example: \textit{"The government took a very important step, but they waited too long for this decision,” Dr. Jose Luis Vargas Segura, a pulmonologist, told Fox News.'} finding \textit{Fox News} as the affiliated ORG. We also sample 100 academic extractions, labeling whether the instance contains RSPEECH, a PERSON and their affiliated university. The accuracy for this is much higher at 87\%.

\subsection{Gender Bias}
36.8\% of extracted speakers have no identifiable gender in gender-guesser. To reduce unknown genders, we take the union of each news outlet's 25 most frequently mentioned people with unknown gender and manually label the gender where the person is recognizable. Most of the names are easily identifiable public figures (e.g., ``Trump'', "Biden", and ``Cuomo''). After this procedure, 26.4\% of extracted sentences have no persons with an identifiable gender.

The majority of androgynous names are Asian names popular both as first and last names. We look at the 25 most frequent names with androgynous labels and manually labeled their gender, if known. We find that the androgynous category captures a unique subset of non-gender-identifying more than androgynous names, so we merge androgynous and unknown gender categories.

\begin{figure}[h!]
\includegraphics[width=\columnwidth]{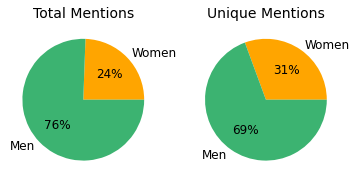}
\caption{\textbf{Gender bias in news.}  Percentage of men and women in all identified expert quotes. We show the composition in total mentions (speakers counted each time they are referenced) and unique mentions (speakers counted once over all mentions). Unique mentions are determined by checking whether each expert's name has a string similarity (via fuzzy string matching) score of 90 or higher to previously mentioned experts. Men are overrepresented in both total and unique mentions. The stronger affinity towards men in total mentions demonstrates that journalists quote the same men repeatedly.}
\label{fig:total_uniq_pies}
\end{figure}

Figure~\ref{fig:total_uniq_pies} breaks down experts quoted in the news by gender. The 26.4\% of instances with unknown gender are omitted to better grasp the immediate disparity between men and women.  The left plot represents the total mentions of all individuals by gender: women represent 24\% of all mentions of experts in the news. 
To identify unique experts, we iterate through all experts while maintaining a list of previously quoted people. For each name, we check whether the person quoted fuzzy string matches to anyone previously quoted with a score of 90 or more. The left pie chart in Fig. \ref{fig:total_uniq_pies} shows the gender breakdown of unique experts, where experts are counted once over all mentions.  Women's representation improves with unique mentions at 31\%. However, this still shows that women are  under-represented in the news, considering that the fields of epidemiology, bio-medicine, and public health---all relevant to the pandemic---have achieved gender parity (or better) \cite{schisterman2017changing,leider2018trends}. Instead, the news media turns to the same group of male experts. The over-representation of men reinforces the idea that science requires traditionally masculine traits and denies fair coverage (and therefore career advancement opportunities) to women.

Sentences quoting men have on average 240 characters per sentence and those quoting women have an average length of 236 characters. This difference is found significant using a two sided t-test (p < 0.01). We also observe that 4.6\% of sentences with expert women also feature an expert man, while only 1.3\% of sentences with an expert man appear with an expert woman.

\begin{figure}[h!]
\includegraphics[width=\columnwidth]{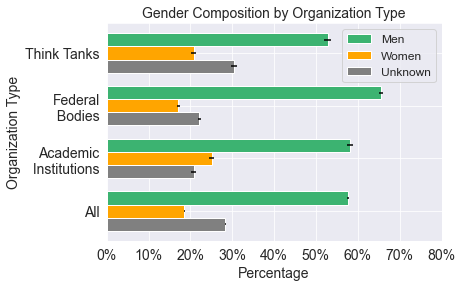}
\caption{\label{fig:orgtype_bs} \textbf{Gender Composition by Organization.} Gender distribution separated by type of organization. Quotes matched to organization types by fuzzy string matching to databases of organization names (Times Higher Educations' 2015 World University Rankings, Index of Federal Departments and Agencies, and On Think Tanks). Error bars determined through bootstrapping 1,000 times. All organization types exhibit gender bias, with federal bodies containing the lowest proportion of women.}
\end{figure}

\subsection{Ideological Bias}

Out of all our extractions, 27.6\% have an organization matching to our academic, federal and think tank databases. Analysis of the organizational breakdown reveals journalists are most likely to reach out to experts affiliated with federal agencies (60.5\%), then academic institutions (21.6\%), and think tanks (17.9\%). One possible explanation is that federal agencies make recommendations for pandemic safety procedures, which are then communicated to the public by reporters.

Fig. \ref{fig:orgtype_bs} shows gender composition by organization type. The bars show average gender representation over 1,000 bootstrapped samples of the data set. The category of unknown gender is included. Experts associated with federal bodies (e.g., CDC, FDA) exhibit the strongest disparity by gender with the lowest percentage of women. Experts from academic institutions manifest less gender disparity, with the highest percentage of women. The lowest percentage of men occurs for experts affiliated with think tanks, which could be due to the high number of persons with ``unknown" gender. 

\begin{figure}[h!]
\includegraphics[width=\columnwidth]{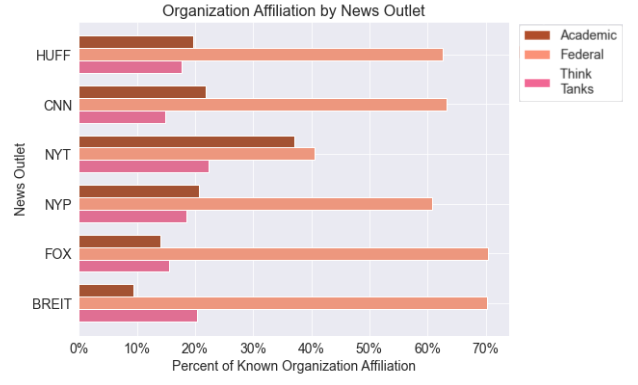}
\caption{\textbf{Preferred Organization Type for Expertise.} 
Distribution of organization types affiliated with news sources in expert quotes. Sources are listed from top to bottom by political leaning reported in Media Bias Fact Check. Across the board, Federal Bodies are the most common type of expertise, though The New York Times has lowest proportion. Breitbart News is the only news outlet with higher use of think tanks than academic institutions.}
\label{fig:orgtype}
\end{figure}

Fig. \ref{fig:orgtype} shows how each news outlet distributes attention over experts from academic institutions, federal bodies and think tanks. Quotes with unknown organization types are not included. We observe that federal bodies are always the most common sources of expertise. NYT quotes federal experts 40.6\%, and all other outlets utilize federal affiliated experts at least 60.8\%. Additionally, we observe that right-leaning outlets typically turn to experts from federal agencies more than left-leaning outlets. Academic institutions are the second most common organization type for experts after federal bodies, except for BREIT and FOX which utilizes academic experts 9.9\% and 14\%, respectively.

\begin{figure}[h!]
\includegraphics[width=\columnwidth]{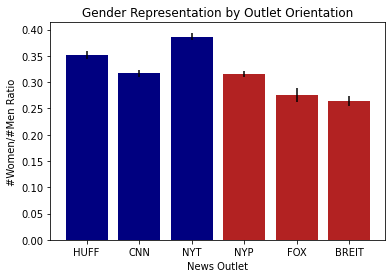}
\caption{\textbf{Ideology and Gender Bias.} Ratio of Women to Men experts quoted by a news source. Smaller ratios signal under-representation of women. Error bars included are from bootstrapping 1000 times. Outlets are ordered left to right by political ideology. Left leaning outlets have the greatest ratio of women cited. The difference in median ratio of news outlets is found significant by the Kruskal-Wallis Test (p $<$ 0.01). }
\label{fig:ratio_stddev}
\end{figure}

Fig. \ref{fig:ratio_stddev} shows gender bias across the ideological spectrum of news outlets, where HUFF, CNN and NYT are classified as liberal (left-leaning) sources, and NYP, FOX, and BREIT as conservative (right-leaning), as reported in Media Bias Fact Check\footnote{https://www.mediabiasfactcheck.com}. The effect of news outlet ideology on gender representation is measured by the ratio of the number of women quoted to the number of men. A ratio of 1.0 signifies equal representation of men and women, smaller ration signal over-representation of men.

All news sources exhibit over-representation of men with ratios at most .387. BREIT has the largest gender disparity with a ratio of 0.264, and NYT has the least gender disparity with the share of women experts at 0.387. 
We use the Kruskal-Wallis H-Test to compare medians for the share of women experts for left-leaning and right-leaning outlets (pictured in blue and red, respectively, in Fig.~\ref{fig:ratio_stddev}). The Kruskal-Wallis test reports a statistic of 8.547 (p $< 0.01$) signifying a statistically significant moderate effect. We conclude left-leaning news outlets exhibit less gender disparity than the right-leaning outlets. 


\subsection{Prestige Bias}

\begin{figure}[h]
\includegraphics[width=\columnwidth]{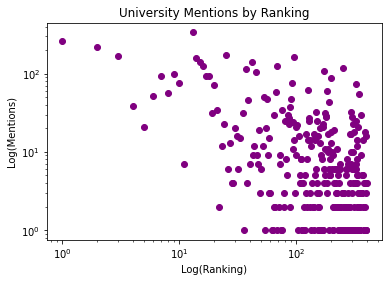}
\caption{\label{fig:loglog} \textbf{Prestige Bias.} Number of mentions of an academic institution in the news as a function of its ranking (for institutions ranked by the Times Higher Educations' World Rankings) shows journalists pay more attention to higher-ranking institutions. Lower rankings signal higher prestige.}
\end{figure}

 We now take a closer look at experts from academic institutions. Fig.~\ref{fig:loglog} shows the number of times an academic institution is mentioned in the news as a function of its placement in the Times Higher Educations' World Rankings. Spearman correlation measures monotonicity between two variables and scores between -1 and 1 (0 means no correlation). The scatter plot shows a downward trend, with a Spearman coefficient of -0.379 (p $<$ 0.01), indicating more prestigious (higher-ranked) institutions generally receive more mentions in the news than less prestigious (lower-ranked) institutions.

We measure prestige bias using the Gini coefficient. Gini is a popular statistical measure of inequality, here attention to academic institutions. A small Gini coefficient means attention (number of mentions of an institution) is equally distributed across universities of any rank, while a Gini coefficient close to one means one university gets all the attention while the rest receive no mentions. The Gini coefficient of mentions of institutions in our data is 0.568, suggesting existence of prestige bias: journalists prefer to turn to experts from the same high-ranking institutions again and again.


\begin{figure}[h]
\includegraphics[width=\columnwidth]{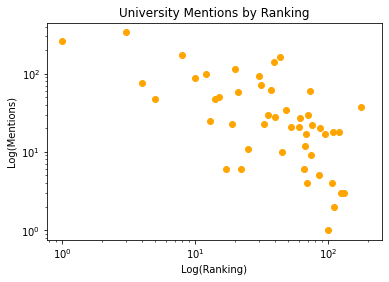}
\caption{\label{fig:healthrank} \textbf{Public Health Ranking and Prestige.} Number of academic institution mentions by public health ranking. In top 48 public health institutions, only a handful with high prestige are heavily utilized by journalists.}
\end{figure}

But what if news outlets are turning to prestige within a domain relevant to the pandemic, like public health? For this case, we rank institutions by prestige in the field of public health using the US News' ranking of US schools of public health\footnote{https://www.usnews.com/best-graduate-schools/top-health-schools/public-health-rankings} in Figure \ref{fig:healthrank}. If journalists were seeking out public health experts, we would expect them to pay more attention to experts from these 48 institutions with higher-ranked schools of public health, resulting in a much lower Gini coefficient. However, the Gini coefficient drops to 0.537, suggesting that prestige bias is driven by extraneous factors such as the institution's ``brand name'' rather than expertise in the relevant field of public health. 


\subsubsection{Ideology and Prestige Bias}

\begin{figure}[h]
\includegraphics[width=\columnwidth]{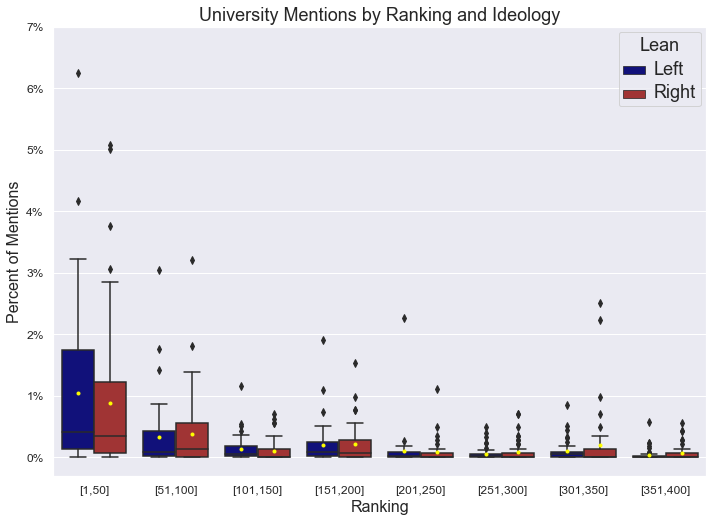}
\caption{\label{fig:leanboxplot} \textbf{Ideology and Prestige Bias.} Boxplot bins the mentions of academic institutions by their rankings, and shows the distributions of the share of mentions of those institutions made by left- and right-leaning news sources. Yellow dots represent group means.
Left-leaning news outlets display stronger preference for experts from prestigious institutions (top-50 ranked universities).}
\end{figure}

We analyze overlap between news outlet ideological leaning and tendency to mention higher ranked universities. The boxplot in Fig. \ref{fig:leanboxplot} shows the distribution of academic expert mentions made by the left-leaning and right-leaning news outlets. The universities which experts are affiliated with are binned by school rank. The boxplot shows the distribution over the share of institution mentions within each bin made by the news sources. 
The boxplot shows the interquartile range, outliers and median for each bin's total mentions. The means within each bin are displayed with yellow points. Prestige bias exists at both ends of the ideological spectrum, though  left-leaning news outlets display more prestige bias, i.e., stronger preference for experts from the top-50 academic institutions. 

We control for political orientation of news outlet in comparing academic institution mentions and rankings. Left-leaning news sources have a Gini coefficient of 0.573 and Spearman coefficient -0.439 (p $<$ 0.01). Right-leaning news sources have a Gini coefficient of 0.562 and Spearman coefficient -0.317 (p $<$ 0.01). This suggests that journalists from conservative sources divide their attention more evenly across institutions than liberal journalists, though the difference is small.

\subsubsection{Gender and Prestige Bias}

\begin{figure}[h]
\includegraphics[width=\columnwidth]{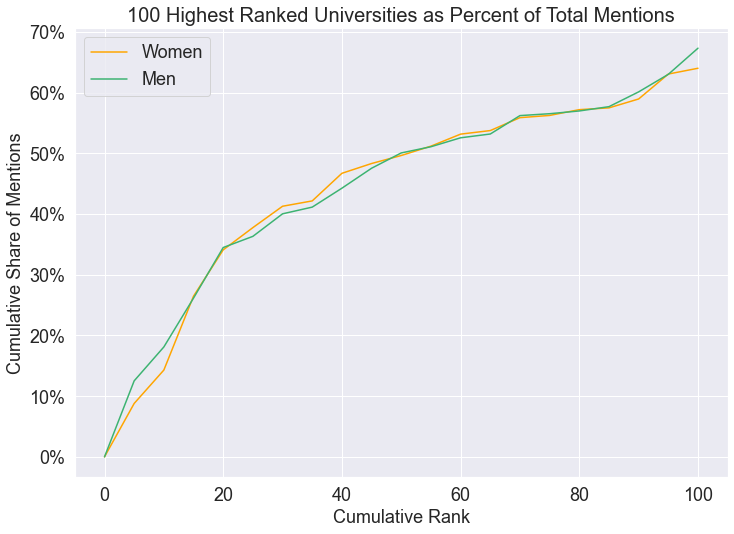}
\caption{\label{fig:prestigegender} \textbf{Gender and Prestige Bias.} Cumulative distribution of mentions for the top 100 institutions broken down by gender. Shows minimal difference in prestige bias between men and women in academia. Roughly one third of quotations come from top 20 institutions, regardless of gender. Men are overrepresented among the quotations from top 10 institutions.}
\end{figure}

Next we examine whether prestige bias varies with expert gender. Fig. ~\ref{fig:prestigegender} shows the cumulative distribution of the share of mentions of experts of either gender affiliated with top-$n$ academic institutions. 
Values of $n$ are 5, 10, 15, etc. We observe almost no difference in how men and women's coverage varies with prestige. For each gender, top-50 highest ranked universities account for half of the academic expert mentions (49.6\% for women and 50.1\% for men). For women, the Gini coefficient of university mentions is 0.56 and Spearman correlation coefficient between the number of mentions and ranking is -.409 (p $<$ 0.01). For men, the Gini coefficient is 0.572 and Spearman coefficient -0.397 (p $<$ 0.01). This disparity shows that prestige inequality is slightly higher for men than women.

We expected that women would need to be from more prestigious institutions to be considered qualified experts. However, we see in Fig. \ref{fig:prestigegender} that there is no significant difference in the prestige distribution for men and women. This lack of difference reveals that gender bias is not substantially amplified within expert mentions from highly ranked universities.


%% file: conclusion.tex
\section{Discussion and Conclusion}

Involving a diverse set of perspectives in the research process enhances quality of research. However, women make up the minority of faculty in most science departments, especially in the more senior and leadership positions \cite{schisterman2017changing}. Additionally, the reward structure of science itself creates disparities through the ``Matthew effect''~\cite{liao2021matthew}, in which highly regarded scientists obtain disproportionate resources and become more likely to produce more successful work. We see this in an example where reviewers in a single-blind peer review process are more likely to accept for publication papers from authors from more prestigious universities \cite{sverdlichenko2022impact}. The researchers from a few prestigious institutions hold a greater influence in shaping scientific research than authors from the less prestigious schools with more diverse populations \cite{morgan2018prestige}. 

Our analysis of a large pandemic-related news corpus shows that women are heard from less frequently than men. Women compose 24\% of expert mentions, though the representation rises to 31\% for unique experts. This suggests that a few men, possibly public figures such as Donald Trump or Andrew Cuomo, are disproportionately represented. Rendering women with less visibility than men paves the way for women's concerns, such as reopening childcare centers and schools, to receive less attention from policy makers.

We observe two different types of ideological bias. The representation of women, measured by the ratio of women included to men, is always higher in left leaning sources than right. Additionally, left leaning news sources display higher prestige bias than right leaning ones. All news sources could improve in representation.

We showed that journalists reporting on Covid-19 paid much more attention to experts with more prestigious affiliations. The gender representation found is a starkly different than that of public health, which is a field one would hope Covid-19 reporting relies upon. When ranking experts by prestige of their institution in the field of public health, ideally the distribution would be somewhat even. However, we observe only a marginally smaller ranking coefficient. This suggests that journalists are either seeking out irrelevant expertise, or wildly misrepresenting the public health field. Journalists have a unique ability to hand pick their subjects, thereby shaping public perception of who constitutes scientific expertise. By focusing their---and the public's---attention on the same small group of high-ranked universities, they risk  perpetuating the cycle of advantage for the privileged minority. To our knowledge, this is the first large scale study of prestige bias in news reporting. 

Our study has a number of limitations. Gender classification is a major limitation. It has been shown that Named Entity Recognition has worse performance identifying women's names as PERSON entities compared to men's names \cite{mehrabi2020man}. As a result, it is likely that our extractions obtained through NER are under-representative of the number of women in the data set. Another gender-based limitation is that the gender predictor used has a misleading androgynous category. Rather than capturing names with equitable gender balance or high association with non-binary people, the androgynous category captures popular Asian last names. The gender classifier is based on a dataset built around cisgender people with historically Western names, meaning our study inherently focuses on cisgender people from Western countries. Such exclusion of non-cisgender people in research continues a long legacy of transgender erasure \cite{keyes2018misgendering}. 

Our work can be expanded by auditing the gender and institutional prestige of Coronavirus experts who are active online on Twitter. We hope to compare network structure by gender category and see how engagement-increasing behaviors differ by gender. We are also interested in hate speech analysis of how scientists of different genders are interacted with on Twitter. Twitter also gives users opportunities to provide their pronouns, allowing us to look at under representations of the gender queer community in scientific research and expert positions.

This large scale analysis of Covid-19 expertise helps us better understand information ecosystems in times of crisis. We observe that men are the dominant sources of expertise, and that a positive feedback loop may occur in news media where men with research success are featured more and therefore are better positioned for further success (and further features in the news media). By automating this analysis, we demonstrate the utility of NLP tools. We hope these findings will help news media more faithfully represent society's diversity.

%% file: ethics.tex
This work uses publicly available published news articles from well known news outlets. Thus, the data set raises few ethical issues around privacy. Ethical concerns around gender inference mechanisms are discussed further in the Conclusion and Discussion portion. The code for this paper will be made available on GitHub.